\documentclass[12pt,a4paper]{article}
\usepackage{graphicx}
\usepackage{times}
\title{\bf Evolution of massive Be and Oe stars at low metallicity  towards
the Long Gamma Ray bursts\footnote{Based on ESO runs 069.D-0275(A), 072.D-0245(A) and (C).}}
\author{C. Martayan$^{1,2}$, J. Zorec$^3$, D. Baade$^4$, Y. Fr\'emat$^5$, \\
J. Fabregat$^6$, and S. Ekstr\"om$^7$\\
\normalsize $^1$ ESO, Alonso de Cordova 3107 Vitacura, Santiago, Chile\\
\normalsize $^2$ GEPI, Observatoire de Paris, place Jules Janssen 92195 Meudon Cedex, France\\
\normalsize $^3$ Institut d’Astrophysique de Paris, 98bis boulevard Arago, 75014 Paris, France\\
\normalsize $^4$ ESO, Karl-Schwarschild-Str. 2, Garching bei Muenchen, Germany\\
\normalsize $^5$ Royal Observatory of Belgium, 3 avenue circulaire, 1180 Brussel, Belgium\\
\normalsize $^6$ Observatorio Astronomico de Valencia, 46980 Paterna Valencia, Spain\\
\normalsize $^7$ Geneva Observatory, University of Geneva, Maillettes 51, 1290 Sauverny, Switzerland
}
\date{\mbox{}}
\begin{document}
\maketitle
\pagestyle{empty}
%
% WE REDEFINE THE plain LaTeX PAGESTYLE !!! 
% THIS PAGESTYLE WILL BE USED FOR THE FIRST PAGE ONLY !
%
\def\bull{\vrule height .9ex width .8ex depth -.1ex}
\makeatletter
\def\ps@plain{\let\@mkboth\gobbletwo
\def\@oddhead{}\def\@oddfoot{\hfil\tiny\bull\quad
``The multi-wavelength view of hot, massive stars''; 39$^{\rm th}$ Li\`ege Int.\ Astroph.\ Coll., 12-16 July 2010 \quad\bull}%
\def\@evenhead{}\let\@evenfoot\@oddfoot}
\makeatother
%
% AND DEFINE OUR MACROS FOR THE REFERENCE LIST
% I.E \beginrefer \refer and \endrefer
%
\def\beginrefer{\section*{References}%
\begin{quotation}\mbox{}\par}
\def\refer#1\par{{\setlength{\parindent}{-\leftmargin}\indent#1\par}}
\def\endrefer{\end{quotation}}
%
% BEGIN THE ABSTRACT CHAPTER WITH \noindent\small, ENCLOSE IT IN A GROUP
% AND BOLDFACE THE TITLE.
%
{\noindent\small{\bf Abstract:} 
Several studies have shown recently that at low metallicity B-type
stars rotate faster than in environments of high metallicity. This is a
typical case in the SMC. As a consequence, it is expected that a larger
number of fast rotators is found in the SMC than in the Galaxy, in
particular a higher fraction of Be/Oe stars.
      Using the ESO-WFI in its slitless mode, the data from the SMC open
clusters were examined and an occurrence of Be stars 3 to 5 times larger
than in the Galaxy was found. The evolution of the angular rotational
velocity at different metallicities seems to be the main key to
understand the specific behavior and evolution of these stars.
      According to the results from this WFI study, the observational
clues obtained from the SMC WR stars and massive stars, and the
theoretical predictions of the characteristics must have the long
gamma-ray burst progenitors, we have identified the low metallicity
massive Be and Oe stars as potential LGRB progenitors. To this end, the
ZAMS rotational velocities of the SMC Be/Oe stars were determined and
compared to models. The expected rates and the numbers of LGRB were then
calculated and compared to the observed ones. Thus, a high probability
was found that low metallicity Be/Oe stars can be LGRB progenitors.
     In this document, we describe the different steps followed in
these studies: determination of the number of Be/Oe stars at different
metallicities, identification of the clues that lead to suppose the low
metallicity Be/Oe stars as LGRB progenitors, comparison of models with
observations.}
%
% NOW COMES THE MAIN BODY OF THE ARTICLE
%
\section{Introduction}
Let us recall that a Be star is a non-supergiant B-type star which spectrum has shown at least once emission-lines.
Actually a Be stars rotate very fast, in the Galaxy at $\sim$ 85\% of the critical rotational velocity.
With episodic matter ejection from the central star, a circumstellar decrection disk is formed.
This phenomenon is not restricted to B type stars but can also occur in late O and early A stars in the Galaxy.
In the first part of this document the metallicity effects on the rotational velocities and on the ratios of Be
to B stars are examined.

The second part of that document deals with the long soft gamma ray bursts (here type 2 bursts) and 
their possible relationship with the massive Be and Oe stars at low metallicity.

\section{Ratios of Be stars with respect to the metallicity}

Maeder, Grebel, \& Mermilliod (1999) found that the ratio of Be stars to B stars in open clusters seems to increase with 
the the metallicity decrease. However, only 1 open cluster in the SMC was used.
Wisniewski \& Bjorkman (2006) found a similar result but the number of open clusters in the SMC/LMC remained small.
The strong variation of the Be/B ratio from an open cluster to another prevents definite conclusions from these studies. 
It was thus needed to increase the samples (much more SMC open clusters) for improving the statistics, 
for better constraining the freedom degrees such as the age, the metallicity, the density, etc, and for quantifying 
the trend found with respect to the metallicity.\\

$\bullet $ \textbf{The WFI H$\alpha$ spectroscopic survey and Be stars ratios}

%\begin{figure}[h]
%\centering
%\includegraphics[width=10cm, height=6cm]{C.Martayan2_fig1.ps}
%\caption{Ratios of Be to B stars per spectral-type categories. 
%Red left bars are for the SMC, right bars for the MW \label{fig1}}
%\end{figure}

Observations with the ESO-WFI in its slitless spectroscopic mode (Baade et al. 1999) were carried out 
on September 25, 2002 with the aim to map the LMC/SMC central parts for detecting 
the H$\alpha$ emission-line stars and the Be stars.
The WFI was used with a grism and a H$\alpha$ filter insuring a bandpass of 700nm, 
a resolving power of 100 adapted to find the emission in the H$\alpha$ line of Be stars. 
The exposure time was 600s. In the SMC 14 images and in the LMC 20 images
were obtained (see Martayan, Baade \& Fabregat 2010).
Let us recall that this kind of instrumentation is not sensitive to the diffuse ambient nebulosity and
is not sensitive enough to the weak emission, thus only  lower estimates of the emission-line stars 
content can be provided.
Finally 3 million spectra were obtained in the SMC (covering 3 square degrees), and 5 million in the LMC.
In this first part of their survey, Martayan, Baade \& Fabregat (2010) focused in 84 SMC open clusters.
Once the emission-line stars and normal stars detected, the stars were classified.

- 1) the astrometry ($\sim$0.5") was performed with the ASTROM code (Wallace \& Gray 2003). 

- 2) the WFI stars were cross-matched with Ogle-II photometric catalogues (Udalski et al. 1998). 
For the stars successfully correlated, the B, V, and I colours were obtained.

- 3) for each open cluster, the E[B-V] and the age were obtained from Pietrzynski \& Udalski (1999), 
then the photometry of the stars was corrected of the reddening.

- 4) the absolute magnitudes of the stars were obtained with the SMC distance modulus (Udalski 2000).

- 5) $\sim$4400 stars in SMC open clusters were classified using the calibration of Lang (2001).

- 6) The SMC data were compared to the results from McSwain \& Gies (2005) in the Galaxy.

%
%In the SMC, the sample is complete down to B5 for normal stars and B3 for the emission-line stars. 
%In the Milky Way (MW), the sample is complete down to A0 for both normal and emission-line stars. 

At that step it is possible to compute the ratios of Be stars to B stars per spectral-type categories, 
i.e. for example B0e/(B0+B0e), etc, in the SMC and MW.
Finally Martayan, Baade \& Fabregat (2010) found that \textbf{the Be stars are 3 to 5 times more abundant 
in the SMC (at low metallicity) than in the MW.}\\

$\bullet $ \textbf{Metallicity (Z) and rotational velocities}

Keller (2004), Martayan et al. (2006, 2007), Hunter et al. (2008) found that OBBe stars rotate faster
at lower Z. The SMC OBBe stars rotate faster than LMC OBBe stars, which rotate faster than their Galactic 
counterparts. This is due to a lower mass-loss (the stellar winds are radiatively driven winds, see Bouret et al. 2003, Vink 2007) 
and lower angular momentum ($\Omega$) loss at lower Z resulting in faster rotation (Maeder \& Meynet 2001). 

The ZAMS rotational velocities for SMC, LMC, and MW Be stars samples were determined by  
Martayan et al. (2007). Be stars at their birth rotate faster 
in the SMC than in the LMC, which rotate faster than in the MW. This is an opacity effect, at lower Z,
the radii should be smaller, thus for an identical intial angular momentum, the stars rotate faster.

Theoretical tracks for Be stars at Z$_{SMC}$ from Ekstr\"om et al. (2008) 
fairly agree with the mean Vsini and mean VZAMS of SMC Be stars (see Martayan et al. 2010).

%#####################################################################################################3

\section{Long soft Gamma Ray Bursts and relationship with Be/Oe stars}
The gamma ray bursts (GRBs) are the most energetic events since the Big Bang.
Among them, 3 classes are distinguished. The type 1 or short GRB (less than 2s) probably resulting of 
the collapse of 2 compact objects. The type 2 or long GRB (LGRB) is associated to the SNIc and is probably
resulting of a massive fast rotating star collapsing in a black hole (Woosley 1993, Fryer 1997). 
And the rare type 3 pseudo LGRB is not associated to a SN.
New theoretical models (Hirschi, Meynet \& Maeder 2005, Yoon, Langer \& Norman 2006, Georgy et al. 2009) provide some informations 
about the LGRBs progenitors masses and rotational velocities from the ZAMS at different Z:
they must be massive fast rotator at low Z (to avoid too much mass-loss and $\Omega$ loss), the H, He envelopes should be lost by fast rotation mixing and/or by a quasi chemically homogeneous evolution 
(Maeder 1987, Yoon, Langer \& Norman 2006; Woosley \& Heger 2006), and the stars should have a weak magnetic field. \\

From the observations point of view, Iwamoto et al. (1998, 2000) found that the massive fast rotating stars 
are at the origin of the LGRBs. Th\"one et al. (2008) found that the LGRB060505 occured in a low-Z galaxy, 
with high star-formation rate, in a young environment (6 Myr) and from an object about 32 M$_{\odot}$.
Campana et al. (2008): the LGRB060218 progenitor had an initial mass of 20 M$_{\odot}$ ($\sim$B0 star), 
with Z = 0.004 ($\sim$ Z$_{(LMC,SMC)}$). Martins et al. (2009) observed several SMC WR stars 
whose evolutionary status and chemical properties can be understood if they are fast rotators 
with a quasi chemically homogeneous evolution.\\

$\bullet $ \textbf{Comparison theory/observations of LGRBs progenitors areas}

\begin{figure}[h!]
\centering
\includegraphics[width=10cm, height=6cm]{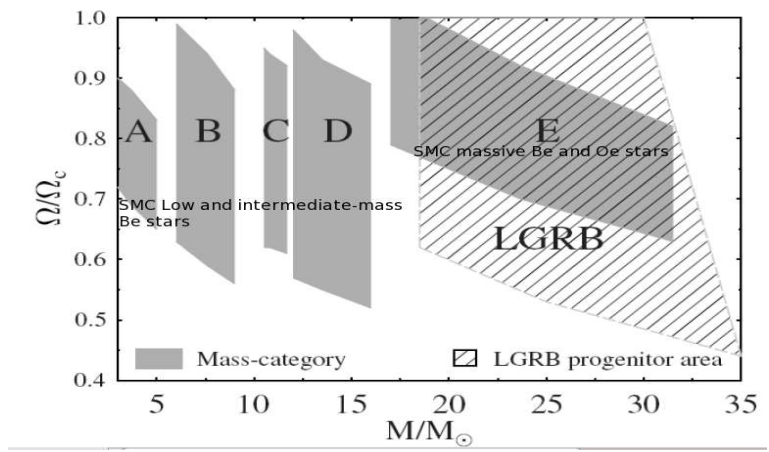}
\caption{Comparison of LGRBs progenitors area (dashed) at Z$_{SMC}$ at the ZAMS 
with categories of SMC Be/Oe stars (grey-plain ABCDE areas). This Figure was adapted from  Martayan et al. (2010). \label{fig2}}
\end{figure}

One can compare the theoretical models  at the ZAMS of Yoon, Langer \& Norman (2006), 
that show areas in rotational velocities vs. the masses for different Z of LGRBs progenitors,
with mass-categories of Be/Oe stars. In Fig.\ \ref{fig2}, such comparison is done using the determined 
ZAMS rotational velocities for Be stars (ABCD) and for massive Be and Oe stars (E) in the SMC by Martayan et al. (2010).

From observational and theoretical clues it seems that massive fast rotators at low Z, i.e, massive Be and Oe stars could be 
progenitors of LGRBs.
The next sections provide additional tests for supporting this finding, including the predictions of the LGRBs rates and numbers in
the local universe.\\

$\bullet $ \textbf{Prediction of the LGRBs rates and numbers from Oe/Be stars populations}

%\begin{table}[h]
%\footnotesize{
%\caption{Predicted LGRBs base rates and rates for given beaming angles depending on the mass category considered. \label{tab1}} 
%\small
%\begin{center} 
%\begin{tabular}{| c | c c c c |}
%\hline 
%Mass-category & LGRBs base-rate & Proba 5$^{o}$ & Proba 10$^{o}$ & Proba 15$^{o}$\\
%\hline
%B1e to O8e & 2.6-3.0x10$^{-4}$ & 1.7-1.9x10$^{-7}$ & 6.7-7.7x10$^{-7}$ & 1.5-1.7x10$^{-6}$\\
%B0e to O8e & 1.7-1.8-x10$^{-4}$ & 1.1-1.2x10$^{-7}$ & 4.4-4.6x10$^{-7}$ & 0.98-1.0x10$^{-6}$\\
%O9e to O8e & 5.3-6.2-x10$^{-5}$ & 3.4-4.0x10$^{-8}$ & 1.4-1.6x10$^{-7}$ & 3.1-3.6x10$^{-7}$\\
%      O8e  & 2.4-2.8-x10$^{-5}$ & 1.6-1.8x10$^{-8}$ & 6.2-7.2x10$^{-8}$ & 1.4-1.6x10$^{-7}$\\
%\hline 
%\end{tabular} 
%\end{center} 
%}
%\end{table} 

To predict the rates of LGRBs from a well defined population, here the massive Be and Oe stars, 
different assumptions and computations were performed:

- the SMC is a representative galaxy of the Im.

- The number of OB stars in the SMC was determined using the SMC OGLE-III catalogue that is complete down to B9.

- The corresponding number of Oe/Be stars is determined with the rates from Martayan, Baade \& Fabregat (2010).

- The binaries were removed of the sample and the transience of the Be phenomenon was taken into account

- The lifetime of different spectral type stars was taken into account.\\

%This provides the base rate of LGRBs, 
%it is then necessary to take into account the beaming angles of the LGRB phenomenon. 
%The predicted rates of LGRBs per angle/galaxy/year are given in Table\ \ref{tab1}.

This provides the base rate of LGRBs, then the LGRBs beaming angles distribution from Watson et al. (2006) is used. 
In such case, the average predicted rate is:  \\
\textbf{R$_{pred}$ LGRBs= (2-5) x 10$^{-7}$ LGRBs/year/galaxy.}

These rates should be compared to the observed value:\\
\textbf{R$_{obs}$ LGRB $\sim$ (0.2-3) x 10$^{-7}$ LGRBs/year/galaxy} 
(Zhang \& M\'esz\'aros 2004; Podsiadlowski et al. 2004; Fryer et al. 2007).\\

%The agreement between the predictions and the observations is relatively good and supports 
%the identification of the LGRBs progenitors.
%Another test is to predict from these rates, the number of LGRBs in the local universe.

%$\bullet $ \textbf{Predicted numbers of LGRBs in the local universe}

One can then predict the number of LGRBs in the local universe, here for a redshift z $\le$ 0.2 
by taking into account that:

- there are 17\% of Im for z$<$0.5 (Rocca-Volmerange et al. 2007),

- the number of galaxies with z $\le$ 0.2 comes from Skrutskie et al. (2006),

- the previous rates are used.

- 11 years are considered from 1998 to 2008, years for which the follow-up of the GRBs is the best 
(in term of redshift and classification) in the GRBOX survey\footnote{http://lyra.berkeley.edu/grbox/grbox.php}.

%The predicted numbers are given in Table\ \ref{tab2} for discrete values of the LGRBs beaming angles.

%\begin{table}[h]
%\footnotesize{
%\caption{Predicted numbers of LGRBs per mass-category in the local universe z $\le$ 0.2 in 11 years and for different beaming angles.
%\label{tab2}} 
%\small
%\begin{center} 
%\begin{tabular}{| c | c c c |}
%\hline 
%Mass category & Number for 5$^{o}$ & Number for 10$^{o}$  & Number for 15$^{o}$ \\
%\hline
%B1e to O8e  & 2-2 & 7-9  & 16-19 \\
%B0e to O8e  & 1-1 & 5-5  & 11-11 \\
%O9e to O8e  & 0-0 & 2-2  & 3-4 \\
%      O8e   & 0-0 & 1-1  & 2-2 \\
%\hline 
%\end{tabular} 
%\end{center} 
%}
%\end{table} 

%However, one can also use the average rates taking into account the distribution of LGRBs beaming angles.
In such case the predicted number is \textbf{N$_{pred} =$ 3-6 LGRBs} in 11 years at z$\le$0.2.

The predicted numbers could be compared to the observed number of LGRBs in 11 years at z$\le$0.2 from the GRBOX survey:
\textbf{N$_{obs} =$ 8 LGRBs}.

From these different tests and comparisons it seems that the populations of low Z massive Be and Oe stars could play 
a major role in the explanation of the LGRBs.

\section{Remaining questions and conclusion}
However, among other there are remaining important questions:

-What is actually the stellar/chemical evolution at low-Z with fast rotation? 
Some discrepancies between the theory and the observations were found by Hunter et al. (2009). 

-The LBV stars could explode in SN without WR phase (Smith et al. 2010), that is not understood by the theory.

- Which observational clue could be found for the chemical evolution? 
Maybe from the GRB environment according to Woosley \& Heger (2006) and Georgy et al. (2009). 
However, the fast rotating stars such as some SMC massive Be/Oe stars have Vsini $\ge$ 500 km/s, 
which according to the models could imply a quasi chemically homogeneous evolution. 

- Is the SMC really representative of all the Im? Are the proportions Be/Oe similar in all SMC-like galaxies? 
Bresolin et al. (2007) observed OB stars in IC1613 (Z$_{IC1613} \le Z_{SMC}$). They found
6 Be stars among the 6 main sequence B stars observed.

- How many LGRBs/SGRBs are not detected by the Gamma Ray observatories?

- There are more LGRBs at high-z. The first stars (very very metal-poor stars) were also probably very fast 
rotators (like Be/Oe stars?) and thus were probably LGRB progenitors.

%
% USE A SECTION WITHOUT NUMBER FOR THE ACKNOWLEDGEMENTS
%
%\section*{Acknowledgements}
%This research is supported in part by contract 007 (MI6). 
%
% BEGIN THE REFERENCE LIST WITH \beginrefer
% USE \refer BEFORE THE REFERENCES AND BEGIN A NEW PARAGRAPH AFTER THE 
% REFERENCE !
% DO NOT FORGET TO END THE LIST WITH \endrefer
%
\footnotesize
\beginrefer

\refer Baade, D., Meisenheimer, K., Iwert, O., et al. 1999, The Messenger, 95, 15

\refer Bouret, J.-C., Lanz, T., Hillier, D. J., et al. 2003, ApJ, 595, 1182

\refer Bresolin, F., Urbaneja, M., Gieren, W., et al. 2007, ApJ, 671, 2028 

\refer Campana, S., Panagia, N., Lazzati, D., et al. 2008, ApJ, 683, L9

\refer Ekstr\"om, , S., Meynet, G., Maeder, A., et al. 2008, A\&A, 478, 467

\refer Fryer, C. L., Mazzali, P. A., Prochaska, J., et al. 2007, PASP, 119, 1211

\refer Fryer, C., Mazzali, P., Prochaska, J., et al. 2007, PASP, 119, 1211

\refer Georgy, C., Meynet, G., Walder, R., et al. 2009, A\&A, 502, 611

\refer Hirschi, R., Meynet, G., \& Maeder, A. 2005, A\&A, 443, 581

\refer Hunter, I., Lennon, D. J., Dufton, P. L., et al. 2008, A\&A, 479, 541

\refer Hunter, I., Brott, I., Langer, N., et al 2009, A\&A, 496, 841

\refer Iwamoto, K., Mazzali, P. A., Nomoto, K., et al. 1998, Nature, 395, 672

\refer Iwamoto, K., Nakamura, T., Nomoto, K., et al. 2000, ApJ, 534, 660

\refer Keller, S. C. 2004, PASA, 21, 310

\refer Lang, K. R. 1992, Astrophysical data, Planets and Stars (New York: Springer Verlag Eds)

\refer Maeder, A. 1987 A\&A, 178, 159

\refer Maeder, A., Grebel, E. K., \& Mermilliod, J.-C. 1999, A\&A, 346, 459

\refer Maeder, A., \& Meynet, G. 2001, A\&A, 373, 555

\refer Martayan, C., Fr\'emat, Y., Hubert, A.-M., et al. 2006, A\&A, 452, 273

\refer Martayan, C., Fr\'emat, Y., Hubert, A.-M., et al. 2007, A\&A, 462, 683

\refer Martayan, C., Baade, D. \& Fabregat, J. 2010, A\&A, 509, A11

\refer Martayan, C., Zorec, J., Fr\'emat, et al. 2010, A\&A, 516, A103

\refer Martins, F., Hillier, D. J., Bouret, J. C., et al. 2009, A\&A, 495, 257

\refer McSwain, M. V., \& Gies, D. R. 2005, ApJS, 161, 118

\refer Pietrzynski, G., \& Udalski, A. 1999, Acta Astron., 49, 157

\refer Podsiadlowski, P., Mazzali, P. A., Nomoto, K., et al. 2004, ApJ, 607, L17

\refer Rocca-Volmerange, B., de Lapparent, V., Seymour, N., et al. 2007, A\&A, 475, 801

\refer Smith, N., Li, W., Filippenko, A., et al. 2010, MNRAS, in press, arXiv1006.3899

\refer Skrutskie, M. F., Cutri, R. M., Stiening, R., et al. 2006, AJ, 131, 1163

\refer Th\"one, C. C., Fynbo, J. P. U., \"Ostlin, G., et al. 2008, ApJ, 676, 1151

\refer Udalski, A., Szymanski, M., Kubiak, M., et al. 1998, Acta Astron., 48, 147

\refer Udalski, A. 2000, Acta Astron., 50, 279

\refer Vink, J. 2007, AIPC, 948, 389

\refer Wallace, P. T., \& Gray, N. 2003, User’s guide of ASTROM

\refer Watson, D., Hjorth, J., Jakobsson, P., et al. 2006, A\&A, 454, L123

\refer Wisniewski, J. P., \& Bjorkman, K. S. 2006, ApJ, 652, 458

\refer Woosley, S.E. 1993, ApJ, 405, 273

\refer Woosley, S. E., \& Heger, A. 2006, ApJ, 637, 914

\refer Yoon, S.-C., Langer, N. \& Norman, C. 2006, A\&A, 460, 199

\refer Zhang, B., \& M\'esz\'aros, P. 2004, Int. J. Mod. Phys. A, 19, 2385

%\refer Zeh, A., Klose, S., \& Kann, D. A. 2006, ApJ, 637, 889

\endrefer  

\end{document}